\documentclass[aps,twocolumn,prl,showpacs,showkeys,floatfix]{revtex4} 
\usepackage{epsfig} 
\usepackage{makeidx} \makeindex 
\usepackage{amsmath,amssymb,amsthm} 
 
\newcommand{\ket}[1]{\,| \, {#1} \,\rangle  \,} 
 
\newcommand{\MyRe}{{\cal R} e \,}   
     
\begin{document}   
\title{Characterization of coherent impurity effects in solid state qubits}

\author{E. Paladino$^{(1)}$, M. Sassetti $^{(2)}$, G. Falci $^{(1)}$ 
 and U. Weiss $^{(3)}$}  
 
\affiliation{ 
$^{(1)}$ MATIS CNR-INFM, Catania  \& 
Dipartimento di Metodologie Fisiche e Chimiche, 
Universit\'a di Catania,
95125 Catania, Italy.\\ 
$^{(2)}$ Dipartimento di Fisica, Universit\`a di Genova \& LAMIA CNR-INFM,
16146 Genova, Italy. \\
$^{(3)}$ II. Institut f\"ur Theoretische Physik,  
Universit\"at Stuttgart, D-70550 Stuttgart, Germany. 
}
 
\begin{abstract}
We propose a characterization of the effects of bistable {\em coherent  impurities} in
solid state qubits. We introduce an effective impurity description in terms of
a tunable spin-boson environment and  solve the dynamics for the qubit coherences.
The dominant rate characterizing the asymptotic time limit is
identified and signatures of non-Gaussian behavior of the quantum 
impurity at intermediate times are pointed out.
An alternative perspective considering the qubit as a measurement device for the  
spin-boson impurity is proposed.
\end{abstract}

 
\date{\today} 
\maketitle 
Coherent nanodevices are inevitably exposed to 
fluctuations due to the solid-state environment. 
Well studied examples are  charged impurities
and stray flux tubes which are sources of telegraphic noise
in a wide class of metallic devices.
Large amplitude low-frequency (mostly $1/f$) noise, 
ubiquitous in amorphous materials~\cite{kn:weissman}, 
is also routinely measured in single-electron-tunneling 
devices~\cite{kn:zorin}. Noise sources are sets of 
impurities located in the oxides and in the substrate, 
each producing a bistable stray polarization. 
Telegraphic noise has also been observed in semiconductor 
and superconductor based nanocircuits~\cite{kn:duty}. 
The possible presence of impurities entangled with the 
device has been suggested in~\cite{kn:simmonds}. 
Recent experiments on Josephson qubits  
indicated that charged impurities may also 
be responsible for noise~\cite{kn:astafiev04} 
exhibiting an ohmic power spectrum at GHz-frequencies.
Different theoretical models have been  
proposed aiming to a unified description of broadband 
noise sources. They share the common idea that the 
variety of observed features 
are due to the dynamics of ensembles of bistable 
impurities~\cite{kn:astafiev04,kn:models,kn:martin,shnirman05}. In particular 
in Ref.~\cite{shnirman05} it has been proposed that  
a noise power 
spectrum compatible with the observed relaxation of 
charge-Josephson qubits~\cite{kn:astafiev04} can be 
obtained if sets of {\em coherent} impurities are considered.

Solid-state noise also determines 
dephasing. This issue has attracted a great deal of 
interest in recent years since it has been recognized as 
a severe hindrance for the implementation of 
quantum hardware in the solid state. The 
effect of slow  noise due to ensembles of 
thermal~\cite{PRL02,kn:galperin} 
and non-thermal~\cite{kn:martin} fluctuators has been 
addressed. Slow noise explains the non-exponential 
suppression of coherent oscillations observed when
repeated measurements are performed~\cite{PRL05,kn:exp}.  
In addition fluctuations active {\em during time evolution}
represent an unavoidable limitation even when 
a single-shot measurement scheme or dynamical 
decoupling protocols~\cite{bang04} are available. 
Note that at experimental temperatures ($\sim 10$ mK)
quantum impurities 
may have a significant influence.

In this Communication we investigate qubit dephasing during time 
evolution due to coupling to a coherent impurity.  
The {\em full} qubit dynamics is solved in the  regime where qubit 
relaxation processes are absent.
We show how the 
coherent and non-linear dynamics of the impurity is reflected in the qubit behavior. 
We identify regimes characterized by strong qubit~-~impurity
back-action. Specifically, we discuss dependence on the impurity preparation and beating
phenomena.
An alternative interpretation with the qubit acting as a measurement device for
the impurity is presented at the end of this Communication.

{\em Model.}--- 
We model the impurity as a two-state system, ${\mathcal H}_{I}= - 
\frac{1}{2} \,\varepsilon\, \tau_z - \frac{1}{2} \,\Delta\, \tau_x $, 
coupled to the qubit ($\sigma$) via 
${\mathcal H}_{QI} = - \frac{1}{2} \,v \,\sigma_z\,\tau_z$ ($\hbar=1$). 
This anisotropic coupling has been discussed for charge 
qubits, where it models the electrostatic interaction~\cite{shnirman05,PRL02,kn:galperin}. 
In this case the two physical states ($\tau_z \to \pm1$) 
correspond to a bistable stray polarization of the qubit.
They are viewed as the ground states of a double-well deformation 
potential, the impurity oscillating coherently between them 
with frequency $\Omega_I = \sqrt{\varepsilon^2+\Delta^2}$.
Dissipative transitions between the minima come from the interaction with
a bosonic bath~\cite{book} 
($\mathcal H_B = \sum_\alpha \omega_\alpha 
a^\dagger_\alpha a_\alpha$) via 
${\mathcal H}_{IB}= - \frac{1}{2} \,\hat{X}\,\tau_z $. 
The operator $ \hat{X} = 
\sum_\alpha \lambda_\alpha (a_\alpha + a^\dagger_\alpha)$
is a collective displacement  with ohmic 
power spectrum   $S(\omega)= 2 \pi \,K  \omega \, 
\, \coth \frac{\omega}{2 T}$ with a high-energy cutoff at $\omega_c$ ($ k_{\rm B}^{} = 1$)~\cite{book}.
This {\em spin-boson environment} (SBE) 
may induce a variety of qubit dynamical behaviors,
since its degree of coherence  depends on $K$ and 
on temperature $T$~\cite{book}.
For instance for  weak damping, $K \ll 1$ 
a crossover occurs between a low ``impurity temperature'', 
$ T \ll \Omega_I$ regime, where 
the impurity performs damped 
oscillations, to the regime of incoherent dynamics 
if $ T \gg \Omega_I$ (white noise  
$S(\omega) \approx 4 \pi K T$)~\cite{chemphys}. 

We assume that the qubit Hamiltonian conserves  
$\sigma_z$, therefore the impurity induces 
{\em pure dephasing}~\cite{book}
with no relaxation of the qubit~\cite{kn:bruder}.
This regime is very interesting since 
energy exchange processes do not blur decoherence of
the qubit, which is then maximally sensitive to the SBE dynamics.
Pure dephasing due to Fano impurities was addressed in~\cite{PRL02}, 
recently the asymptotic dynamics has been studied~\cite{kn:lerner}.
This model corresponds to a 
over-damped impurity (SBE at $K=\frac{1}{2}$), here
we consider $K \ll 1$ where the impurity may behave coherently.  

\begin{figure}[t!]
\resizebox{80mm}{28mm}{\includegraphics{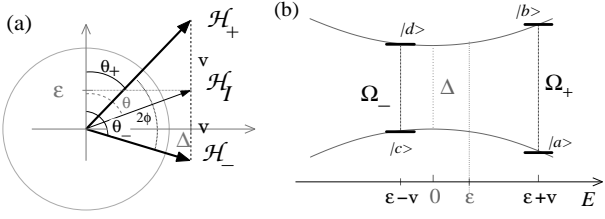}} 
\caption{
(a) Impurity Bloch sphere. An isolated impurity
 ${\cal H}_I$ defines the mixing angle  
$\theta = \arctan \Delta/\varepsilon$,   
${\mathcal H}_\pm$ define 
$\theta_\pm = \arctan \Delta/(\varepsilon \pm v)$. 
(b) Impurity bands $\pm \sqrt{E^2 +\Delta^2}$: 
impurity energy splittings depend on the qubit state, 
$\Omega_{\pm}= \sqrt{(\varepsilon \pm v)^2 + \Delta^2}$.
Eigenstates of ${\mathcal H}_0$, 
are $\{ \ket{i} \}$, $i \,=\, a,\, b,\, c,\, d$.
Conservation of $\sigma_z$ allows only intra-doublet processes, $a \leftrightarrow b$, $c \leftrightarrow d$.
\label{fig:splittings}}
\end{figure}

{\em Method and analytic results.}--- 
For pure dephasing the qubit Hamiltonian can be gauged 
away by a proper rotation. In this picture 
we consider the reduced density matrix 
$\rho(t)$ 
describing the entangled qubit-impurity system. For 
$K \ll 1$ the
interaction with the bosonic bath is studied by the
Born-Markov master equation (ME)~\cite{kn:cohen} 
\begin{eqnarray} 
\partial_t  \rho(t) 
&&\hskip-9pt=  - i [{\mathcal H}_0, \rho(t)]  - \hskip-2pt  \int_0^
\infty  \hskip-7pt d t^\prime  
 \Big\{\mbox{$\frac{1}{4}$} \,S( t^\prime)\, [\tau_z, \, 
[\tau_z( t^\prime), \rho(t) ]] 
\quad\nonumber\\  
&&\hskip-9pt+ \, \mbox{$\frac{i}{2}$} \,  
\chi(t^\prime)\, [\tau_z, \, [\tau_z( t^\prime), \rho(t) ]_{\mbox{\tiny$+$}}]
\Big\} \, , 
\label{eq:ME} 
\end{eqnarray} 
where $\mathcal H_0= {\mathcal H}_{QI} +  {\mathcal H}_{I}$ is the undamped Hamiltonian. 
Here, the transform $S(t)$ of the power spectrum and 
the bath susceptibility $\chi(t) = -i \langle
[\hat{X}(t), \hat{X}(0)]_+ \rangle \Theta(t)$
enter the damping term.
We introduce the {\em conditional} Hamiltonians of the impurity
${\mathcal H}_\pm = - \frac{1}{2} (\varepsilon \pm v) \, \tau_z - \frac{1}{2} \Delta \,
\tau_x \,$, (see Fig.~\ref{fig:splittings}) and the eigenvectors of 
${\mathcal H}_0$, $\{ \ket{i} \}$, which are factorized
in eigenstates of $\sigma_z$ and of ${\mathcal H}_\pm$~\cite{chemphys}.
The qubit dynamics at pure dephasing is described by the coherences
$\langle \sigma_\pm(t) \rangle = {\rm Tr}[\rho(t)\,(\sigma_x \pm i \sigma_y)
\otimes \openone_\tau]$, 
and in particular 
$$
\langle \sigma_-(t) \rangle=  
 \,  
\big[ \rho_{ac}(t) + \rho_{bd}(t) \big]\cos \phi +  
 \, 
\big[ \rho_{ad}(t) - \rho_{bc}(t) \big] \sin \phi \, , 
$$
where 
$\phi = \frac{1}{2}(\theta_- - \theta_+)$ 
is a combination of the mixing angles of 
${\mathcal H}_\pm$ (Fig.~\ref{fig:splittings}).
Since $\sigma_z$ is conserved, the damping 
tensor presents only four non vanishing $4 \times 4$ 
diagonal blocks.  
We focus on the block acting on the terms entering  $\langle
\sigma_-(t) \rangle$.  Performing a partial
secular approximation within this block,
we get two sets of decoupled equations for
$\rho_{ac},\rho_{bd}$ and  $\rho_{ad},\rho_{bc}$. We quote here the
first set
\begin{eqnarray} 
\left(\hskip-3pt\begin{array}{c} 
\dot{\rho}_{ac}(t) 
\rule[-7pt]{0pt}{1pt} \\ 
\dot{\rho}_{bd}(t) 
\end{array} 
\hskip-3pt\right) 
\,=\,  
\left(\hskip-3pt 
\begin{array}{cc} 
i \delta   - \Gamma_1  
&  
 \Gamma_{12}  
\rule[-7pt]{0pt}{1pt}\\ 
  \Gamma_{21}  & 
-i \delta  - \Gamma_2  
\end{array} 
\hskip-3pt\right)  
\left(\hskip-3pt\begin{array}{c} 
\rho_{ac}(t) \rule[-10pt]{0pt}{1pt}\\ 
\rho_{bd}(t) 
\end{array} 
\hskip-3pt\right)\hskip-3pt \, ,
\label{eq:ME-sec1} 
\end{eqnarray} 
where $\delta=\frac{1}{2}(\Omega_+ - \Omega_-)$,
Fig.~\ref{fig:splittings}. 
The rates $\Gamma_i$, describing dissipative
transitions and pure dephasing processes between the $4$-states and the bosonic bath, read 
\begin{eqnarray} 
\hskip-2pt 
\begin{array}{ll} 
\Gamma_{1,2} &\hskip-5pt = 
\alpha_{+}^2 \Gamma_{\mp}(\Omega_{+}) + \alpha_{-}^2
\Gamma_{\mp}(\Omega_{-})
+  \eta_s \,S(0) \, , 
\rule[-7pt]{0pt}{1pt} \\ 
 \Gamma_{12,21} &\hskip-5pt= \alpha_{+} \alpha_{-}
\, 
[  \Gamma_{\pm}(\Omega_+) +  \Gamma_{\pm}(\Omega_-) ]  \, ,
\rule[-7pt]{0pt}{1pt} \\
\alpha_{\pm} &\hskip-5pt= \frac{1}{2\sqrt 2} \sin \theta_\pm;\;
\eta_s =
\frac{1}{2} \, \sin^2\!\bar{\theta} \,\sin^2\! \phi \,,
\end{array} 
\label{rates}
\end{eqnarray}
where $\bar{\theta}= \frac{1}{2} (\theta_++\theta_-)$. Here 
$\Gamma_\pm(\omega)= 2\pi K \omega[\coth(\frac{\omega}{2T}) \pm 1]$, 
are the impurity emission $(+)$ and absorption $(-)$ rates of energy $\omega$. 
The elements $\rho_{ad},\rho_{bc}$ satisfy similar equations with 
$\delta$ replaced by 
$\Omega = \frac{1}{2}(\Omega_{+}+\Omega_{-})$ and rates 
\begin{eqnarray*} 
\hskip-2pt 
\begin{array}{ll} 
\Gamma_{3,4} &\hskip-5pt= \alpha_{+}^2 \Gamma_{\mp}(\Omega_+) +  
\alpha_{-}^2 \Gamma_{\pm}(\Omega_-)  
        +  \eta_c\, S(0) \, , 
\rule[-7pt]{0pt}{1pt} \\ 
\Gamma_{34,43}&\hskip-5pt =  \alpha_{+}\alpha_{-}\, 
                [  \Gamma_{\pm}(\Omega_+)+ \Gamma_{\mp}(\Omega_-) ]  \, , 
\rule[-7pt]{0pt}{1pt} \\ 
\eta_c &\hskip-5pt =
\frac{1}{2} \, \cos^2\!\bar{\theta} \,\cos^2\! \phi \, . 
\end{array} 
\end{eqnarray*} 
Diagonalization of Eq. (\ref{eq:ME-sec1}) and of the corresponding 
set for $\rho_{ad},\rho_{bc}$   
yields the eigenvalues 
\begin{eqnarray} 
\hskip-2pt 
\begin{array}{ll} 
\lambda_{1,2} &\hskip-5pt=   - 
\frac{\Gamma_1 + \Gamma_2}{2}
\pm 
 \frac{1}{2} \sqrt{(2 i \delta +\Gamma_2 -\Gamma_1)^2 + 4 
\Gamma_{12}\Gamma_{21}} \, , 
\rule[-7pt]{0pt}{1pt} \\ 
\lambda_{3,4}  &\hskip-5pt = - 
\frac{\Gamma_3 + \Gamma_4}{2}
\pm  
\frac{1}{2} \sqrt{(2 i \Omega +\Gamma_4 -\Gamma_3)^2 + 4 \Gamma_{34}
\Gamma_{43}} \, . 
\label{lambdaME} 
\end{array} 
\end{eqnarray}
The explicit form of $\langle \sigma_-(t) \rangle$ depends on the 
initial conditions for $\rho(t)$. 
Because of the high accuracy of preparation presently achieved
in solid state implementations, factorized qubit-impurity states
$\rho(0) = \rho_{\sigma}(0) \otimes \rho_{\tau}(0)$, represent a realistic
scenario. The impurity initial state is instead out of the experimentalist 
control, thus
we choose
$\rho_\tau(0) = \frac{1}{2} \,( \openone_\tau \, + \,  p_z \,\tau_z)$, $p_z$
being the initial average of $\tau_z$. 
The  impurity starts from a totally unpolarized state for $p_z=0$,
from a pure state if $p_z=\pm 1$. 
This class of initial states guarantees the
positivity of the dynamical process ensuing from Eq.(\ref{eq:ME}).
With this choice we find
\begin{eqnarray} 
\langle \sigma_-(t) \rangle &=& {\langle \sigma_- (0)\rangle} \,\, 
{\textstyle \sum_{i}} \,A_{i}e^{\lambda_i t} \, , 
\label{eq:sigmatime} 
\end{eqnarray} 
\vspace{-8mm} 
\begin{eqnarray} 
\hskip-2pt 
\begin{array}{ll} 
A_{1,2}&\hskip-5pt = 
 \frac{\cos \!\phi }{2(\lambda_1-\lambda_2)}\left\{\cos \!\phi\, 
[\lambda_1-\lambda_2\pm(\Gamma_{12} + \Gamma_{21})] \mp \right.  
\rule[-10pt]{0pt}{2pt} \\ 
&\hskip-5pt \left. p_z\cos(\theta + \bar\theta)[-2i\delta 
-\Gamma_2 +\Gamma_1 +\Gamma_{12}  
-\Gamma_{21}]\right\} \, , 
\rule[-10pt]{0pt}{2pt} \\ 
A_{3,4} 
&\hskip-5pt  = \frac{\sin\!\phi }{2(\lambda_3-\lambda_4)}\left\{\sin\!\phi\, 
[\lambda_3-\lambda_4\mp(\Gamma_{34} + \Gamma_{43}) ]\pm\right.  
\rule[-10pt]{0pt}{2pt} \\ 
&\hskip-5pt  \left. p_z\sin(\theta + \bar\theta) 
[-2i\Omega -\Gamma_4 +\Gamma_3 -\Gamma_{34} +\Gamma_{43}]\right\}  \,.
\label{weight} 
\end{array} 
\end{eqnarray} 
Eqs.~(\ref{lambdaME})-(\ref{weight}) are
the main result of this Communication. 
They cover the 
parameters regime where $S(\Omega_\pm) \ll \Omega_\pm$. 
Single-phonon processes dominate at low $T$,
whereas 
multiphonon-exchanges are paramount at higher $T$ where the
white noise results 
of 
\cite{chemphys} are recovered.
Reliability of ME is confirmed by a real-time path-integral 
calculation.

{\em Discussion of the results.}--- 
We focus our analysis on the low-temperature regime $T \ll \Omega_{-}$.
Here effects of the dissipative processes internal to the SBE on the qubit behavior 
are clearly identifiable. 
In this limit energy absorption processes are 
exponentially suppressed ($\Gamma_-(\Omega_\pm)\approx 0$)
and the eigenvalues take the forms
\begin{eqnarray} 
\hskip-2pt 
\begin{array}{ll} 
\lambda_1 &\hskip-5pt = i \delta - \eta_s S(0) \, , 
\rule[-10pt]{0pt}{2pt} \\  
\lambda_2 &\hskip-5pt = -i \delta -  
       \frac{\displaystyle \gamma_{r+} + \gamma_{r+}^0 +\gamma_{r-} + 
\gamma_{r-}^0}{4}
        - \eta_s S(0) \, , 
\rule[-10pt]{0pt}{2pt} \\ 
\,\lambda_{3,4} &\hskip-5pt =  \, \pm \, i \Omega   
        -  \frac{\displaystyle\gamma_{r\mp} + \gamma_{r\mp}^0}{4} 
        - \eta_c S(0) \, ,
\end{array} 
\label{eq:poleslowT} 
\end{eqnarray} 
where  intra-doublet relaxation rates  (see Fig.~\ref{fig:splittings}) 
\begin{equation} 
\gamma_{r\pm} = \frac{1}{2}\sin^2(\theta_\pm) S(\Omega_\pm)= 
\frac{1}{2} \Big(\frac{\Delta}{\Omega_\pm}\Big)^2 S(\Omega_\pm)
\label{relaxation} 
\end{equation} 
have been introduced ($\gamma_{r\pm}^0$ value at $T=0$). Note that pure dephasing
processes $\propto S(0)$, are not simple sum  of
intra-doublet dephasing terms, $\gamma_{\phi\pm} =
\frac{1}{2}\cos^2(\theta_\pm) S(0)$.  
\begin{figure}[t!]
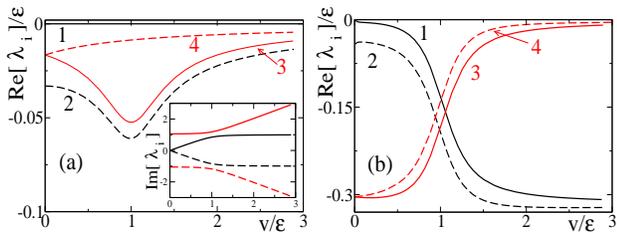
 
\centering 
\resizebox{40mm}{30mm}{\includegraphics{figure2a}}
\resizebox{40mm}{30mm}{\includegraphics{figure2b}}
\caption{The four rates $\MyRe[\lambda_{i}]$ from Eqs.(\ref{lambdaME}). 
In (a) $T=0$. Inset:
  imaginary parts (independent on temperature for $T < \Omega_+$).
  In (b) $T=0.5 \Delta$.  Parameters are $K=0.1$, 
$\varepsilon=3 \Delta$.
}
\label{fig:eigen} 
\end{figure}  

In the following we present a selection of illustrative behaviors for $\varepsilon > \Delta$.
In this regime the two conditional Hamiltonians ${\mathcal H}_{\pm}$ may differ significantly
and enforce peculiar impurity dynamical behaviors.
For example beatings when $\delta$ approaches 
$\Omega$, i.e. around $\varepsilon=v$ which 
identifies a sort of  ``resonance regime'' for our problem. 

We first characterize the asymptotic qubit dynamics,
by the $T$ and $v$ dependence of the eigenvalues.
At zero temperature the pure dephasing contributions fade away, and
one rate, $\MyRe[\lambda_1]$, 
vanishes, as expected.  Only
emission processes contribute to the residual rates, and they directly
sound out intra-doublet relaxation rates $\gamma_{r\pm}^0$.
Their behaviors reflect the sensitivity of 
${\mathcal H}_\pm$, to noise acting  
along $\tau_z$. While $\gamma_{r+}^0$ decreases with increasing $v$, $\gamma_{r-}^0$ takes a  
maximum at the resonance point (see Eq.(\ref{relaxation})), 
the ``transverse'' ($\theta_-=\pi/2$) noise condition for ${\mathcal H}_-$.  
This implies a non-monotonous dependence of $\MyRe[\lambda_{2,3}]$ on the coupling 
$v$, Fig.~\ref{fig:eigen}(a).  The  imaginary parts of
$\lambda_{1,2}$ and  $\lambda_{3,4}$ interchange characters at 
resonance (Fig.~\ref{fig:eigen}(a) inset) leading to possible
hybridization (see below).
Increasing $T$ leading correction to the rates come from  pure dephasing 
terms $S(0)$.
As a difference with $T=0$, all the rates  are finite and cross around resonance,
Fig.~\ref{fig:eigen}(b).
 
These features are  crucial for 
the asymptotic dynamics of $\langle \sigma_-(t)\rangle$,
which does not depend on the impurity preparation.
We then expect at $T=0$, undamped oscillations with
$\delta$, while at finite $T$, damped oscillations driven by
one or two complex eigenvalues.  For example 
in the case of  Fig.~\ref{fig:eigen}(b) the dominant rate
is $\MyRe[\lambda_1]$ for $v<\epsilon$ and $\MyRe[\lambda_4]$ for $v>\epsilon$.
It is a non-monotonous function
of $v$ and a cusp signals crossing of
eigenvalues (a similar effect may explain non-monotonic behavior
of~\cite{kn:lerner}).
\begin{figure}[t!]
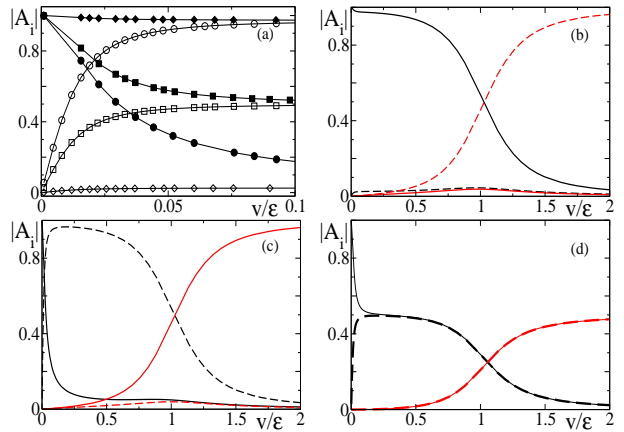
 
\centering 
\resizebox{40mm}{28mm}{\includegraphics{figure3a}}
\resizebox{40mm}{28mm}{\includegraphics{figure3b}}
\resizebox{40mm}{28mm}{\includegraphics{figure3c}}
\resizebox{40mm}{28mm}{\includegraphics{figure3d}}
\caption{Weights $|A_i|$ of $e^{\lambda_it}$ from Eq.(\ref{weight})  
as a function of $v/\varepsilon$.
(a) Dominant weights in the small $v$ region: $|A_1|$ (full symbols) and 
$|A_2|$ (open symbols) for  $p_z=0$ (squares), $p_z=-1$ (circles) and $p_z=1$ (diamonds).
Effect of impurity preparations:  $p_z=1$ (b),
 $p_z=-1$ (c), and  $p_z=0$ (unpolarized state) (d).
$|A_{1}|$ (blue), $|A_{2}|$ (blue dashed), $|A_{3}|$  (red), 
$|A_{4}|$ (red dashed).
Parameters: $T=0$, $\varepsilon=3\Delta$, $K=0.1$.} 
\label{fig:weight} 
\end{figure} 

At intermediate times, all eigenvalues may be relevant, depending on
the weights $A_i$ in Eq.(\ref{weight}).  To substantiate this point,
 in Fig.~\ref{fig:weight} we show $A_i$ corresponding to the eigenvalues
in  Fig.~\ref{fig:eigen}(a) for different 
preparations. Remarkably, the weights very weakly depend on $T$ (not shown),
then the following picture generally holds for $T< \Omega_-$.
For extreme weak coupling, $v \to 0$, $|A_1|\approx 1$ 
(Fig.~\ref{fig:weight}(a)) implying 
universal dynamics independent on the initial conditions.
The dominant eigenvalue is $\lambda_1$ with 
$\delta\to 0$ and $\langle \sigma_-(t) \rangle$ decays
exponentially with the  Golden rule rate
$\Gamma_{GR} = \frac{v^2}{2}\frac{S(0)}{\varepsilon^2+\Delta^2}\sin^4
\theta$.  In this regime the impurity acts as a Gaussian reservoir and
may be described with linear response theory in the coupling $v$.  
Away from this tiny region non-Gaussian effects occur  
and different impurity preparations result in different time behaviors,  
giving  separate information on the various eigenvalues. 
Far from resonance, a single frequency  
shows up in $\langle \sigma_-(t) \rangle$  independently on $p_z$ 
($\delta$ if $v < \varepsilon$, $\Omega$ if $v > \varepsilon$). 
Damping of the oscillations depends on the initial condition,
Fig.~\ref{fig:weight} (b)~-~(d). 
For instance, at finite $v < \varepsilon$, the decay occurs with 
$\MyRe[\lambda_{1}]$ if  $p_z=1$ and with $\MyRe[\lambda_{2}]$ if  $p_z=-1$, 
both rates are present for unpolarized initial state. 
This behavior is stable against temperature variations.
Beatings {\em and} $T$~-~dependence are instead characteristic of the resonant regime.
At $v=\varepsilon$, at least two amplitudes are equal,
$|A_1| \approx |A_4|$ ($p_z=1$) or $|A_2| \approx |A_3|$ ($p_z=-1$).
Damped beatings at  $\Omega_{\pm}= \Omega\pm\delta$ are
possible due to the hybridization of 
$\Omega\approx\delta$ (Fig.~\ref{fig:eigen}(a) inset).

\begin{figure}[t]
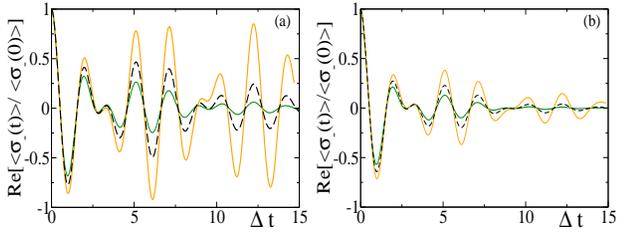
 
\centering 
\resizebox{40mm}{30mm}{
\includegraphics{figure4a}}
\resizebox{40mm}{30mm}{
\includegraphics{figure4b}}
\caption{ $\MyRe[\langle \sigma_-(t) \rangle/ 
\langle\sigma_-(0)\rangle]$  at resonance  $\varepsilon=v=3\Delta$
for initial states (a) $p_z=1$: slow decay with $\MyRe[\lambda_{1/4}]$,
(b) $p_z=-1$: fast decay with  $\MyRe[\lambda_{2/3}]$. Parameters:
$T=0$ (blue), $T=0.5 \Delta$ (red), $T=0.9 \Delta$ (green), $K=0.1$.} 
\label{fig:resigmatime} 
\end{figure}

We illustrate these features  in  Fig.~\ref{fig:resigmatime} for  $v=\varepsilon$. 
The beatings visibility is reduced with increasing $T$, due the onset
of the pure dephasing processes.
For an  unpolarized state, $p_z=0$,   $\langle \sigma_-(t) \rangle$ shows 
a intermediate behavior between the ones at $p_z=\pm 1$ since 
at resonance all eigenvalues contribute (Fig.~\ref{fig:weight}(d)).
Damping is strongest for  $p_z=-1$,
weakest for  $p_z=1$ and intermediate for $p_z=0$.
In fact, for $\epsilon > \Delta$, preparation 
in the pure state $p_z = +1$ makes the impurity close to its ground
state and less damped, while it is close to the excited state when $p_z= -1$ with 
strongest damping.

\begin{figure}[b!]
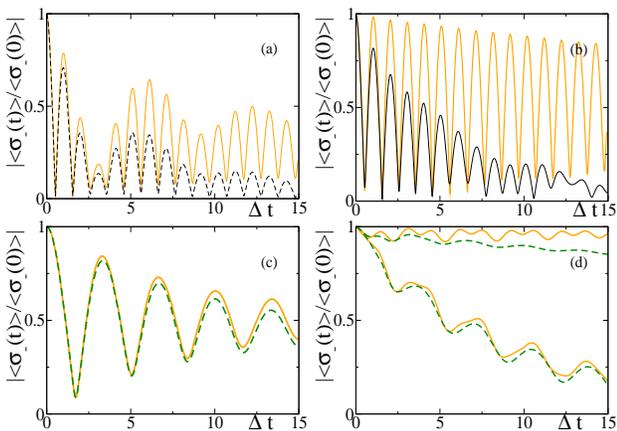
 
\centering 
\resizebox{40mm}{28mm}{\includegraphics{figure5a}}
\resizebox{40mm}{28mm}{\includegraphics{figure5b}}
\resizebox{40mm}{28mm}{\includegraphics{figure5c}}
\resizebox{40mm}{28mm}{\includegraphics{figure5d}}
\caption{$|\langle \sigma_-(t)\rangle/\langle\sigma_-(0)\rangle|$  
for $\varepsilon= 3\Delta$, $K=0.1$. 
Panels (a) and (b): resonant impurity $v=\varepsilon$.
(a)  $p_z=0$ at  $T=0$ (black) and $T=0.5 \Delta$ (red);
(b) $T=0$ for $p_z=1$ (orange), $p_z=-1$ (black). 
Panels (c) and (d): non resonant case  $v= \Delta$ at $T=0$ (blue) and  
at $T=0.9 \Delta$ (red). 
In (c) $p_z=0$, in (d) $p_z=1$ top, $p_z=-1$ bottom. Note the weak $T$-dependence.
} 
\label{fig:modulosigmatime1} 
\end{figure}

In the last part of this Communication we present an alternative perspective, 
considering the qubit as a measuring device for a mesoscopic system 
described by the SBE.
Remarkably, the qubit acts as a detector despite the absence of direct qubit-SBE
inelastic transitions~\cite{kn:aguado}. 
In fact, the pure dephasing coupling
amounts to a ``dispersive'', quantum non-demolition measurement regime for the qubit.
Detection is feasible due to the qubit back-action on the SBE.
This point of view is illustrated in Fig.~\ref{fig:modulosigmatime1}, 
where the length of the Bloch vector in the
${\hat x}$~-~${\hat y}$ plane, $|\langle \sigma_-(t)\rangle|$,
acts as  a sensitive detector of the mesoscopic system (``impurity'') 
preparation. 
At resonance, the unpolarized state, $p_z=0$, is identified  by
beatings, Fig.~\ref{fig:modulosigmatime1}(a).
These almost disappear for pure states, 
$p_z=\pm 1$, where oscillations  at $\Omega_+$ occur,
Fig.~\ref{fig:modulosigmatime1}(b).
Identification of the impurity preparation far from resonace results instead from  
different oscillation amplitudes and/or
decay rates, Fig.~\ref{fig:modulosigmatime1} (c)~-~(d).  

In conclusion, we have identified in time domain non Gaussian and back-action effects
due to a coherent bistable impurity.
These may represent a ultimate limitation for solid state qubits even 
when single shot measurement schemes are available. 
Our analysis by changing temperature,  
strain $\varepsilon$ and coupling $v$, 
may provide valuable insights to realistic scenarios where a wide distribution 
of the parameters has to be considered~\cite{kn:galperin}.
The employed  SBE  represents a general effective model for complex physical
baths awaiting specific microscopic description, as those
typical of solid state nanodevices.

We acknowledge support from the EU-EuroSQIP (IST-3-015708-IP) and MIUR-PRIN2005 
(2005022977).

\end{document}